# Beam test calibration of the balloon-borne imaging calorimeter for the CREAM experiment

P. S. Marrocchesi[c], H. S. Ahn[b], M. G. Bagliesi[c], A. Basti[d], G. Bigongiari[c], A. Castellina[e], M. A. Ciocci[c], A. Di Virgilio[d], T. Lomtatze[d], O. Ganel[b], K. C. Kim[b], M. H. Lee[b], F. Ligabue[d], L. Lutz[b], P. Maestro[c], A. Malinine[b], M. Meucci, [c], V. Millucci, [c], F. Morsani, [d], E. S. Seo[a,b], R. Sina[b], J. Wu[b], Y. S. Yoon[a], R. Zei[c], S.-Y. Zinn[b]

*(a) Dept. of Physics, University of Maryland, College Park, MD 20742 USA*
*(b) Inst. for Phys. Sci. and Tech., University of Maryland, College Park, MD 20742 USA*
*(c) Dept. of Physics, University of Siena and INFN, Via Roma 56, 53100 Siena, Italy*
*(d) INFN sez. di Pisa and Scuola Normale Superiore , 2 v. F.Buonarroti, I-56127 Pisa, Italy*
*(e) IFSI sez. di Torino and INFN, 4 Corso Fiume 4, 10133 Torino, Italy*
Presenter: Pier Simone Marrocchesi (marrocchesi@pi.infn.it), ita-marrochesi-P-abs2-og15-poster

CREAM (Cosmic Ray Energetics And Mass) is a multi-flight balloon mission designed to collect direct data on the elemental composition and individual energy spectra of cosmic rays. Two instrument suites have been built to be flown alternately on a yearly base. The tungsten/Sci-Fi imaging calorimeter for the second flight, scheduled for December 2005, was calibrated with electron and proton beams at CERN. A calibration procedure based on the study of the longitudinal shower profile is described and preliminary results of the beam test are presented.

## 1. Introduction

CREAM (Cosmic Ray Energetics And Mass) is a balloon experiment for direct measurements of cosmic ray composition and energy spectra, designed to approach the PeV scale in a series of flights taking advantage of the new Ultra Long Duration Balloon (ULDB) concept under development by NASA. The instrument was flown for the first time in December 2004 with a conventional LDB balloon from Antarctica. A new flight duration record of about 42 days was established and about 40 million of science events were collected.
In order to allow for annual flights, two instruments suites (herein called CREAM-1 and CREAM-2) are planned to be flown on alternate years. In this way, the refurbishment operations that follow the recovery of one payload can take place almost simultaneously with the flight preparation of the second payload. The CREAM-1 instrument [1, 2, 3] includes a sampling tungsten/scintillating fiber calorimeter preceded by a graphite target with scintillating fiber hodoscopes, a pixelated silicon charge detector (SCD), a transition radiation detector (TRD) and a segmented timing-based particle-charge detector (TCD).
A second flight is scheduled for December 2005 from McMurdo with the CREAM-2 instrument configuration. The latter does not include the TRD, while the charge identification performances of SCD are enhanced by the addition of a second layer of pixelated silicon sensors. The calorimeter to be flown on the second payload was calibrated with high-energy electrons and protons in the H2 beam line at CERN in September 2004. Its construction and performance are described in more detail in [4].
In this paper, we report on a method to equalize the response of the 1000 calorimeter cells, based on the study of the average longitudinal shower development at different electron beam energies.

## 2. The structure of the CREAM calorimeter

An inelastic interaction of the primary nucleus in the densified graphite target ($\sim 0.5\ \lambda_{int}$) initiates a hadronic shower with a narrow electromagnetic core, generated by the decay of neutral pions, which is imaged by a 20 $X_0$ (radiation length) calorimeter with an active area of $50{\times}50$ cm$^2$. The Tungsten/Sci-Fi stack is made of twenty tungsten plates with interleaved active layers instrumented with 1 cm wide ribbons of 0.5 mm diameter scintillating fibers. The light from each ribbon is fed, via an acrylic light-mixer, into a bundle of clear fibers and then split into 3 sub-bundles, each connected to one pixel of a Hybrid PhotoDiode (HPD). This scheme allows to divide the calorimeter dynamic range into 3 sub-ranges of different gain, optimized to match the dynamic range of the front-end electronics. For mechanical reasons, alternate ribbons are read out on opposite ends, therefore one calorimeter layer (50 ribbons) is read-out in two half-layers, each comprised of 25 ribbons and connected to one HPD unit. A total of 2560 channels are read out from 40 photodetectors arranged in 4 crates.

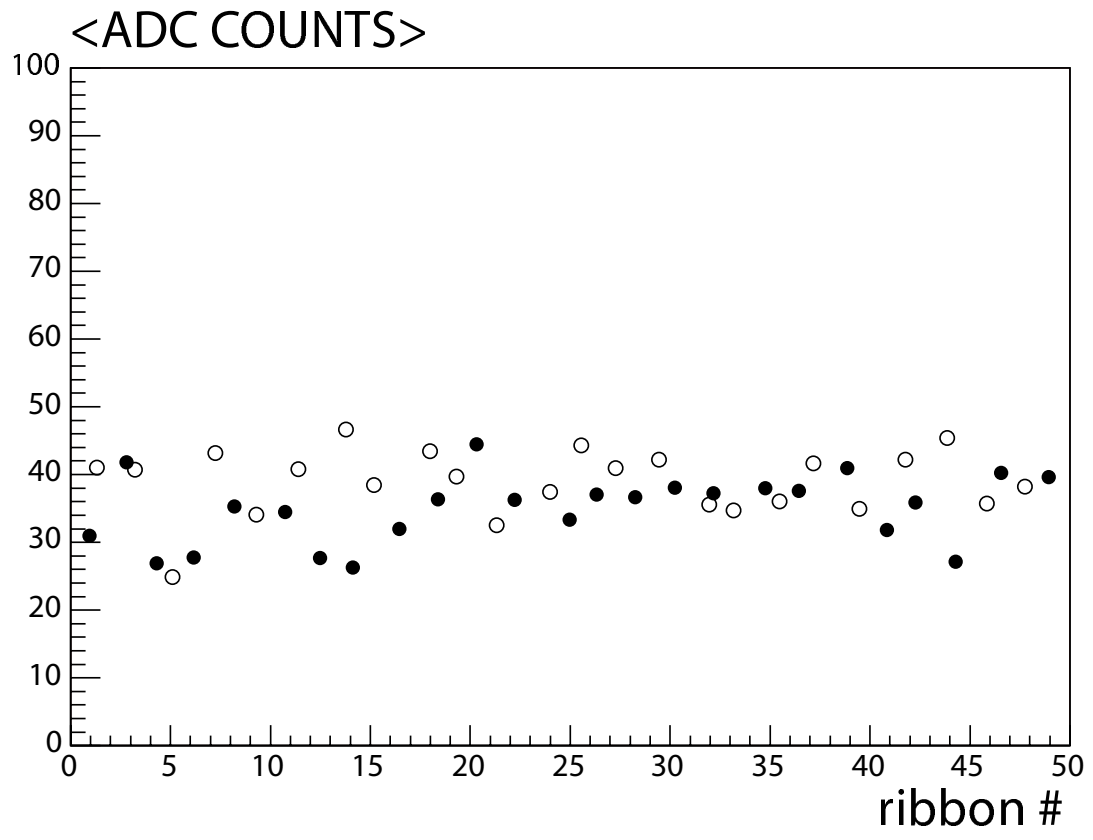

**Figure 1.** Average response of odd (filled circles) and even (open circles) ribbons in a calorimeter layer during the calorimeter beam scan.

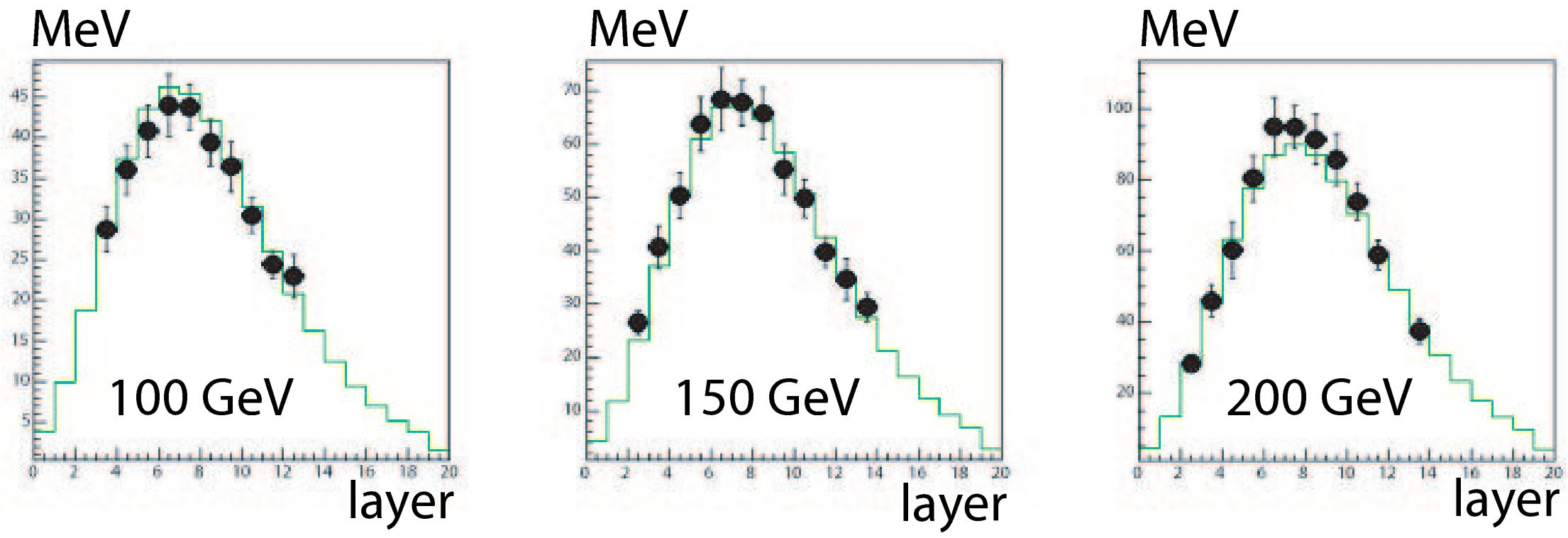

**Figure 2.** Average longitudinal profiles with electrons of energy 100, 150, 200 GeV. The solid line is the MonteCarlo prediction.

## 3. Calibration data sets

A first set of electron data was collected by steering 150 GeV electrons onto the center of each ribbon and scanning the calorimeter in both views. A second set of electron data at the same energy was taken with a Pb absorber (∼5 $X_0$) in front of the target. The resulting shift in the position of the shower maximum allowed the calibration of the relative response of the first layers of the calorimeter. A third data set was taken by rotating the calorimeter by 180 degrees. In this configuration, the beam impinged on the bottom layer of the calorimeter and a tungsten absorber of thickness 2 $X_0$ was added in front of the beam to compensate both for the absence of the target (1 $X_0$) and for the smaller thickness, in terms of g/cm$^2$, of the bottom Al plate with respect to a standard tungsten plate. Addition of a further 5 $X_0$ Pb absorber resulted in a longitudinal distribution with shower maximum on layer 16, which allowed calibration of the last 4 layers of the calorimeter. Data were also taken with proton beams of 150, 250, 350 GeV at different angles. Proton data results will be published separately.

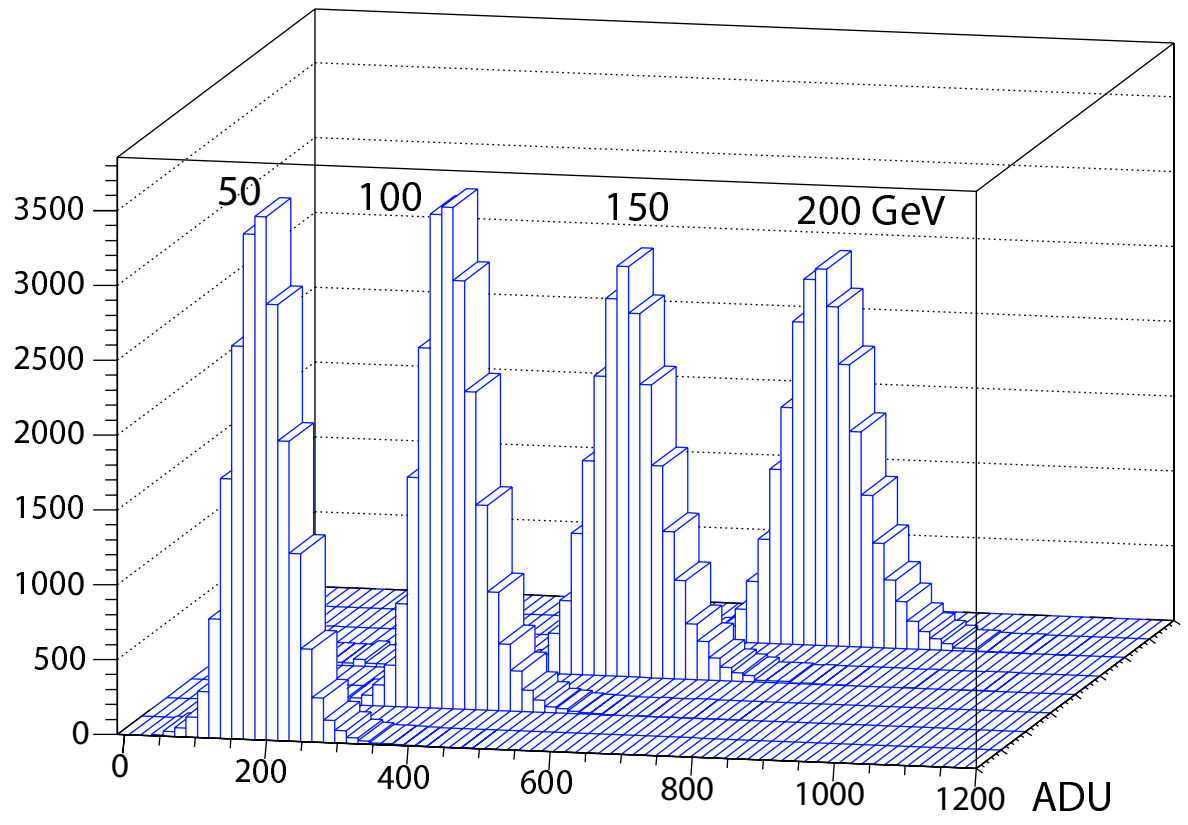

**Figure 3.** Total calorimetric energy with electron beams of 50, 100, 150, 200 GeV in ADC units.

## 4. Channel equalization and longitudinal profile deconvolution

Data collected during the horizontal and vertical beam scans, with 150 GeV electrons, were studied to equalize the calorimeter at channel level. A two steps procedure was implemented.
First, the response of each ribbon connected to a same HPD (averaged over about 5000 events) was obtained. Then, each group of 25 ribbon signals was equalized to the average response of the respective half-layer (herein referred to as "average HPD signal"). Since the gain uniformity of the HPD pixels is known to be at the level of a few % , an observed variance of order 10% (for an example, see Fig.1) is expected to have contributions from the accuracy in the calorimeter construction (including individual light yield and optical coupling efficiency of the ribbons), but also from systematic errors on the beam position accuracy during the scan.

The second step of the procedure consisted in equalizing the 40 HPDs average signals. This was needed in order to take in account the differences in quantum efficiency (QE) and optical coupling among the photodetectors. This correction is not straightforward because the scintillation light collected by each HPD is a function of its depth in the Tungsten/Sci-Fi stack at a given energy. Therefore, the HPDs average responses have to be first corrected for the dependence on the longitudinal shower profile. This "deconvolution method" was applied only at a one beam energy (150 GeV) where, by comparing the longitudinal shower profile obtained with the 40 HPDs average signals with the one predicted by a Monte Carlo simulation (based on the Fluka 2003.1b package [6]), a set of 40 energy independent constants, one per HPD, was extracted. The same procedure was repeated with data sets taken at different energies, as a consistency check of the energy independence of the HPD constants.

As a result of the two step procedure, 1000 calibration constants (one per ribbon) were determined, using exclusively the 150 GeV electron data. Application of the calibration constants to the data taken at different beam energies resulted in average longitudinal profiles as in Fig.2, where the solid line is the MonteCarlo prediction. The data points were found to be consistent with the simulation.

## 5. Energy calibration

During the energy scan, the calorimeter was kept at fixed position, with the beam impinging on the centre at normal incidence. The calorimetric energy was calculated from the sum of the 3 ribbons centered on the beam, as shown in Fig.3 for the four different beam energies. This procedure removes the contribution of the noise due to the calorimeter cells far away from the shower envelope and reproduces the effect of the sparsification threshold, which is applied to the calorimeter data during the flight in order to reduce the overall size of the event. The calorimeter response to electron energy up to 200 GeV is shown in Fig.4(left) where the intercept from the linear fit was found to be zero within errors and the slope is consistent with a sampling fraction of 0.53% as predicted by the simulations.
The dependence of the calorimetric measurement with the applied voltage (HV) of the photodectors has been studied separately, by taking data at different HV values in the range 6 to 10.5 KV at a fixed beam energy. The calorimeter response was found to be linear in the above range of HV values and the high voltage scan has been repeated at different beam energies. The energy resolution dependence on the HV gain of the HPD is shown in Fig.4(right) for electrons at a beam energy of 150 GeV (upper curve) and 200 GeV (lower curve).

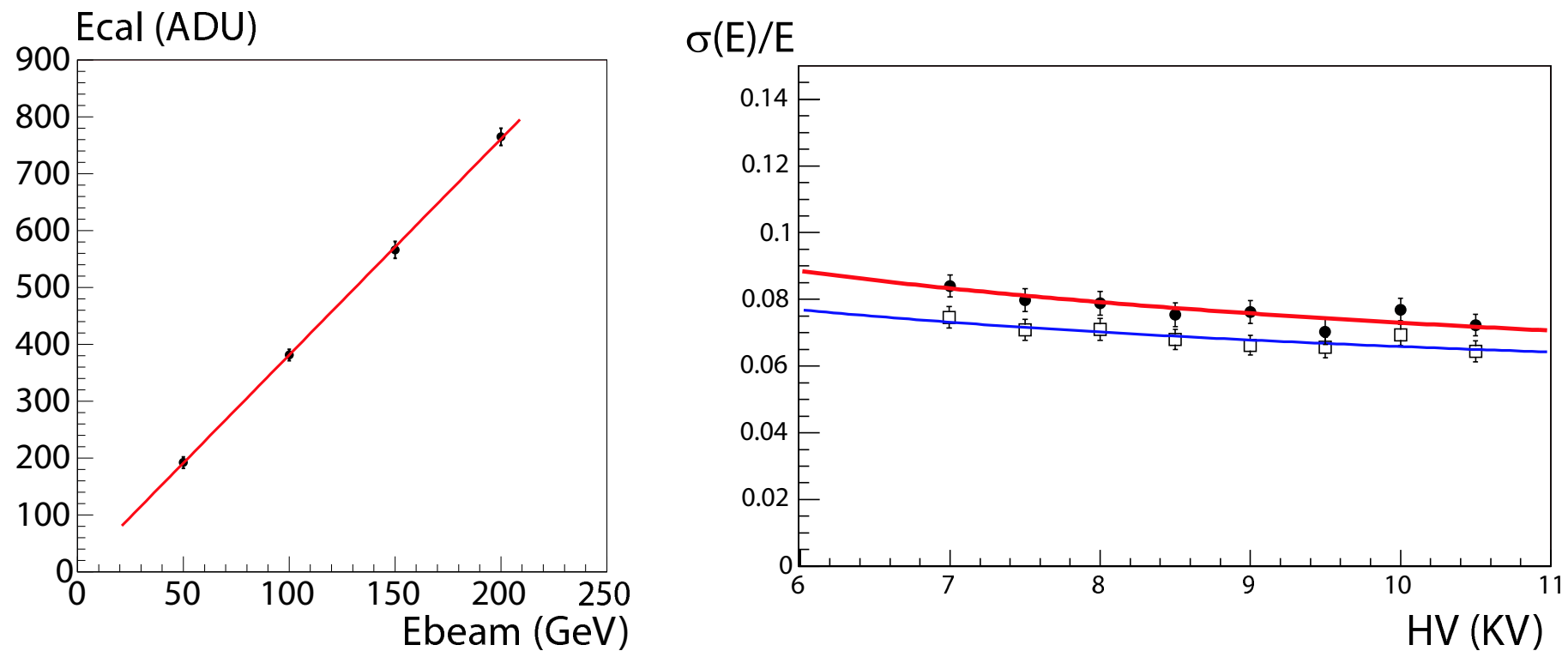

**Figure 4.** (left) Calorimetric energy (in ADC units) vs. beam energy (GeV) for electron runs; (right) Energy resolution vs. photodetector high voltage for electrons at 150 (upper curve) and 200 GeV (lower curve).

## 6. Acknowledgements

This work was supported by NASA grants in the US and by INFN in Italy. The authors greatly appreciated the support of CERN for the beam test facilities and operations.